# A controllable nanomechanical memory element

Robert L. Badzey[a)], Guiti Zolfagharkhani[a)], Alexei Gaidarzhy[b)] & Pritiraj Mohanty[a)]

a) Department of Physics, Boston University, Boston, Massachusetts 02215

b) Department of Aerospace and Mechanical Engineering, Boston University, Boston, Massachusetts 02215

We report the realization of a completely controllable high-speed nanomechanical memory element fabricated from single-crystal silicon wafers. This element consists of a doubly-clamped suspended nanomechanical beam structure, which can be made to switch controllably between two stable and distinct states at a single frequency in the megahertz range. Because of their sub-micron size and high normal-mode frequencies, these nanomechanical memory elements offer the potential to rival the current state-of-the-art electronic data storage and processing.

Since the comprehensive design of the first analytical engine by Babbage[1] in 1834, the concept of storage and manipulation of data in a mechanical state has continued to shape the evolution of modern computers. The development of the electronic transistor[2] and magnetic storage methods[3,4] heralded the coming of technologies that outstripped the mechanical approach both in manipulation speed and data density. However, recent advances in the fabrication and measurement of nanomechanical systems enable the realization of mechanical computation of a size and speed comparable to those of electronic or magnetic systems[5]. The fundamental requirement of computational memory is that it can store and manipulate information. This has traditionally been realized through the binary element, which has two distinct states designated 0 and 1. Elements are either non-volatile, retaining their state information without the need for an external stimulus, or volatile, resetting themselves once the stimulus is removed. Fidelity is crucial for all memory elements, volatile and non-volatile alike. The relationship between the 0 and 1 state and the characteristics of each state need to be stable. Additionally, speed of manipulation is essential for volatile elements. Therefore, elements must perform their functions at speeds and densities comparable to the current standards.

We present a doubly-clamped bistable nanomechanical beam as the basis for our potential memory element. Although bistable mechanical beams have been known since the time of Euler[6,7,8], their use as bona fide memory elements have been daunted by the lack of control of the two Euler states in real time. In addition, the normal mode frequencies of macroscopic beams are generally quite low. However, recent work on single-wall nanotube (SWNT) systems[5] has demonstrated the feasibility of a nanoscale mechanical memory element. The nanoscale realization of the mechanical memory element is exciting because of its frequency of operation. In principle, these devices can be operated in the MHz - GHz range, with densities exceeding the superparamagnetic limit for magnetic disk drives (100 GB/in$^2$).

Our initial wafer is a Si/SiO2/Si heterostructure in which all the Si layers are single crystals. E-beam lithography is used to realize the beam and pad design. Following development, 85 nm of Au with a 10 nm Cr underlayer were evaporated to define the beam, electrically connect it, and serve as a mask for the etching process. The final step is a combination of two separate RIE processes, one anisotropic to define the sidewalls of the beam, and one isotropic to undercut the structure and suspend it. This sample has dimensions of 8 μm x 300 nm x 200 nm, yielding a resonant frequency of 23.568 MHz. Samples are wirebonded and placed inside a $^3$He cryostat and cool to 275 mK at the center of a 9 T superconducting solenoid magnet. We excite the normal modes of the beam using a magnetomotive technique[9,10]. Using either a network analyzer or RF source, we drive a MHz-frequency current through the contact electrode; the perpendicular magnetic field is 8 T. The interaction between the magnetic field and oscillating current creates a Lorentz force *F=ILB*, where I is the excitation current, L is the beam length and B is the magnetic field. This force is perpendicular to both the field and the current, and results in a transverse

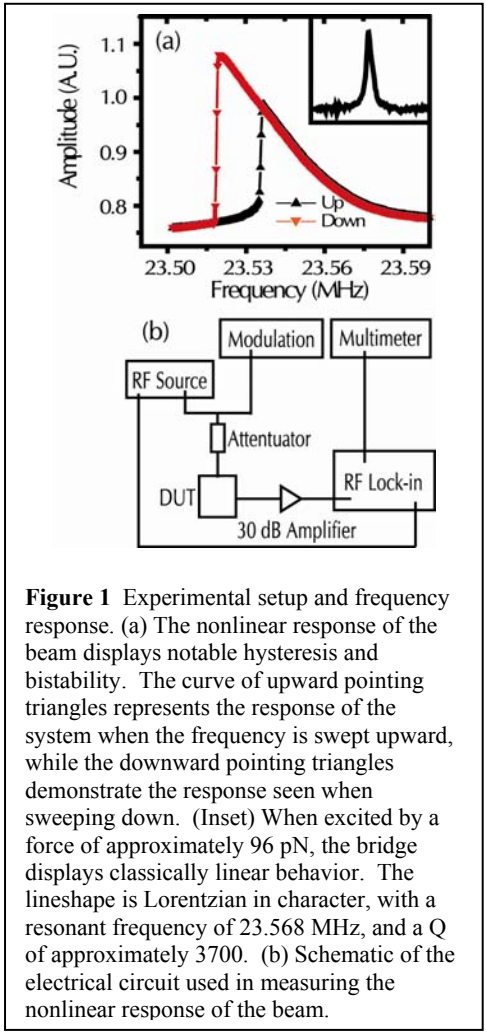

**Figure 1** Experimental setup and frequency response. (a) The nonlinear response of the beam displays notable hysteresis and bistability. The curve of upward pointing triangles represents the response of the system when the frequency is swept upward, while the downward pointing triangles demonstrate the response seen when sweeping down. (Inset) When excited by a force of approximately 96 pN, the bridge displays classically linear behavior. The lineshape is Lorentzian in character, with a resonant frequency of 23.568 MHz, and a Q of approximately 3700. (b) Schematic of the electrical circuit used in measuring the nonlinear response of the beam.

oscillation in the beam. The induced EMF due to this motion is proportional to the beam displacement via

$$\Delta x = V_{EMF}/\xi LB\omega \qquad (1)$$

where $\xi$ is an integration constant which depends on mode shape. This technique is sensitive enough to easily detect displacements less than 1 Å[11].

Figure 1a shows the characteristic nonlinear response of the beam, obtained with a vector network analyzer. The beam response demonstrates a notable hysteresis, creating a range of frequencies in which the beam is bistable[12]. The inset shows the linear response of the beam; the Lorentzian lineshape corresponds to that of a damped driven harmonic oscillator with a Q ~ 3700 and a linear resonance frequency of 23.568 MHz. When nonlinear response is achieved, such an oscillator is described by the well-known Duffing equation:

$$\ddot{x} + 2\gamma\dot{x} + \omega_0^2 x \pm k_3 x^3 = F\cos\omega t. \qquad (2)$$

It is important to note that the resonant frequency of the beam displays a negative frequency shift with thermal cycling and increasing drive amplitude[8]. In addition, the voltage difference between the two bistable states is dependent on the location of the excitation frequency within the bistable region. The sharp drop on the left-hand side of the response curve is the signature of a beam which undergoes mode softening when excited nonlinearly[13]. Using the measurement circuit described in Figure 1b, we drive the beam nonlinearly at a single frequency within the bistable region and modulate it with a second frequency. The modified Duffing equation incorporating the modulation reads

$$\ddot{x} + 2\gamma\dot{x} + \omega_0 x \pm k_3 x^3 = F\cos\omega_d t + F_{mod}(\omega_{mod}), \qquad (3)$$

and the square-wave modulation has the form

$$F_{mod}(\omega_{mod}) = A_{mod}\Theta(T); \quad \Theta(T)\begin{cases} 0 \to (n-1)T < t < (n-\frac{1}{2})T \\ 1 \to (n-\frac{1}{2})T < t < nT \end{cases}. \qquad (4)$$

The excitation frequency $\omega_d$ is 23.497 MHz, with an $A_{excite}$ of 4.0 dBm, equivalent to an RMS voltage of 350 mV. Our modulation is 0.05 Hz with $A_{mod}$ between -0.3 dBm and 1.0 dBm. These amplitudes are taken at the instrument ports; the signal is attenuated significantly before reaching the top of the fridge, and even more before reaching the actual sample. The signal from the sample is amplified by 30 dB on its way to an RF lockin amplifier. We find that a switch event shows up as a 60 µV jump in the signal. We can factor out the power amplification and find that the actual signal size is equal to $60\mu V/\sqrt{10^3} \approx 60\mu V/31 \approx 2\mu V$. Using Equation 1, this means that a switch event corresponds to a displacement of 4 Å.

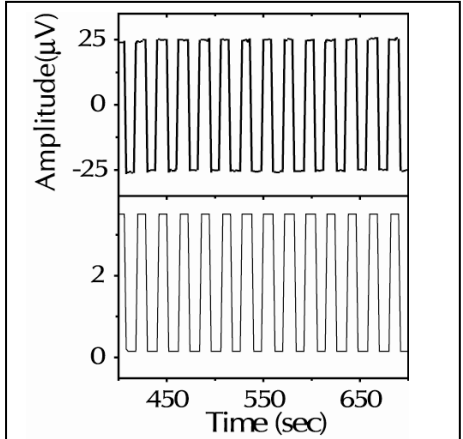

**Figure 2** Observation of switching events. A lock-in measurement of the beam at f = 23.497 MHz (upper graph) shows distinct two-state switching in response to a square wave modulation (lower graph). The frequency of modulation is 0.05 Hz, and the amplitude is 1.0 dBm. It is important to note that the phase and period of the beam response match those of the modulation, and that the amplitude of the switches corresponds to the size of the nonlinear jump.

Using the low-frequency modulation, we are able to coherently control the behavior of the beam, switching it between states at will. Figure 2 shows a representative switching sequences, with the attendant modulation signal, taken at 275 mK with an $A_{mod}$ of 1.0 dBm. Note that the response of the beam tracks both the frequency and the phase of the modulation signal.

The two states correspond in voltage to the upper and lower extremes of the hysteresis curves shown in Figure 1a, and are independent of the actual modulation power (Figure 3). At low $A_{mod}$, the beam responds by skipping periods, but the switches that do occur have the same amplitude. Increasing $A_{mod}$ results in the beam hitting more and more available switches, until a threshold power $A_C$ is reached and full switching is established. Again, the modulation signal is presented for comparison; the period of the modulation signal is matched, even when periods are skipped. Once a potential switching event is skipped, a subsequent switch will not occur until the modulation signal goes through at least one full period. The graph at the bottom shows the relative population of the upper and lower states as a function of modulation power. At lower modulation powers, there is a tendency to remain in the lower state. Below a threshold value, the modulation is no longer strong enough to induce switching, and the beam remains in either the upper or lower state indefinitely.

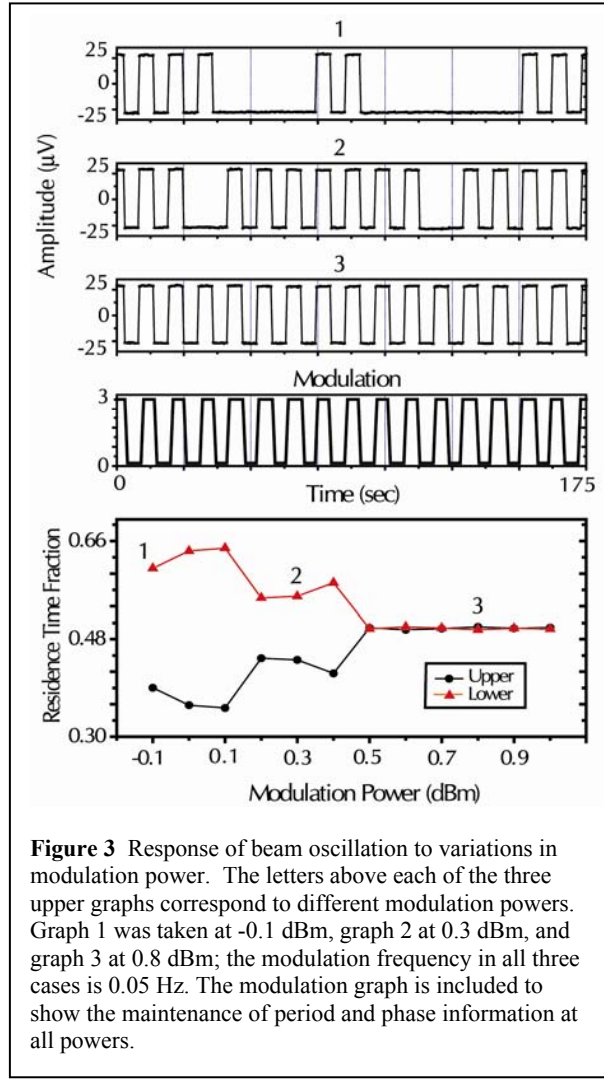

**Figure 3** Response of beam oscillation to variations in modulation power. The letters above each of the three upper graphs correspond to different modulation powers. Graph 1 was taken at -0.1 dBm, graph 2 at 0.3 dBm, and graph 3 at 0.8 dBm; the modulation frequency in all three cases is 0.05 Hz. The modulation graph is included to show the maintenance of period and phase information at all powers.

The high normal-mode frequencies for these mechanical structures arise because of their small size, as $f_0 \propto \dfrac{t}{L^2}$ [14,15]. Therefore, while an 8 μm-long beam yields a resonance frequency ~23.5 MHz, a beam 1 μm-long has a fundamental mode above 1 GHz[16]. While we excite the beams at much lower frequencies, the open-loop limit to modulation frequency is $\omega_0/Q$. We are currently exploring measurement techniques which will allow us to track modulations up to the resonant frequency of the beam.

Not only does the small size help in increasing the operation frequency, it offers three fundamental advantages for applications in data storage and manipulation. First, the areal density also goes up as the second power of the size, $1/L^2$.

Nanomechanical memory elements can be arranged in densities that exceed the superparamagnetic limit (100GB/in$^2$). The first demonstration of data storage beyond this limit involves the so-called millipede with an array of micron-sized mechanical cantilevers operating at a kilohertz speed[17,18]. The second fundamental advantage is the level of power consumption in the read-write process of actuation and sensing. The composite signal from all sources required to excite nonlinear behaviour and induce switch events was measured to be less than -48 dBm (900 μV) at the top of the fridge. The third advantage is the resilience of the mechanical structure to electrical and magnetic fields.

While several mechanical schemes exist, our method is attractive for several reasons. First, as this scheme does not require physical contact of our devices we avoid stiction, which is the combination sticking/friction effect prevalent in micro- and nanomechanical systems[19]. Additionally, by using the dynamics of the beam itself, we achieve switching with no additional processing or external readout elements. While the single-wall nanotube (SWNT) scheme[5] shows promise, it also requires as-yet-undemonstrated control over the fabrication and placement of the SWNTs. Full fabrication control of NEMS devices is commonplace, in placement and element uniformity.

In conclusion, we have observed and controlled coherent switching of a nanomechanical memory element by modulating the nonlinear bistable modes of a micron-size Si beam. With a linear normal mode frequency of 23.56 MHz, this beam represents a significant step forward in the possible realization of fast and robust mechanical memory elements.

We acknowledge the NSF (Nanoscale Exploratory Research (NER) grant number ECS-0404206), and the DOD/ARL (DAAD 19-00-2-0004). We also acknowledge partial support from the Sloan Foundation and the NSF (DMR-0346707, CCF-0432089 and ECS-0210752).